\documentstyle[12pt]{article}
\def\ie{{\em i.e.}}
\def\eg{{\em e.g.}}

\def\beq{\begin{equation}}
\def\eeq{\end{equation}}

\catcode`\@=11 
\def\coeff#1#2{{\textstyle{#1\over #2}}}

\def\VEV#1{\left\langle #1\right\rangle}
\def\vev#1{\langle #1\rangle}
\def\lsim{\mathrel{\mathpalette\@versim<}}
\def\gsim{\mathrel{\mathpalette\@versim>}}
\def\@versim#1#2{\vcenter{\offinterlineskip
    \ialign{$\m@th#1\hfil##\hfil$\crcr#2\crcr\sim\crcr } }}

\def\JL{J. L. Lopez}
\def\DVN{D. V. Nanopoulos}

\def\r#1{$\bf#1$}
\def\rb#1{$\bf\overline{#1}$}

\def\t1{{\tilde 1}}

\def\GeV{\,{\rm GeV}}

\def\to{\rightarrow}

\def\NPB#1#2#3{Nucl. Phys. B {\bf#1} (19#2) #3}
\def\PLB#1#2#3{Phys. Lett. B {\bf#1} (19#2) #3}

\begin{document}
\begin{flushright}
\baselineskip=12pt
CTP-TAMU-37/95\\
DOE/ER/40717--16\\
ACT-12/95\\
\tt hep-ph/9510216
\end{flushright}

\begin{center}
\vglue 0.5cm
{\Large\bf Vanishing ${\rm Str}\,{\cal M}^2$ in the presence of
anomalous~$\rm U_A(1)$\\}
\vglue 0.5cm
{\Large Jorge L. Lopez$^1$ and D.V. Nanopoulos$^{2,3}$\\}
\vglue 0.25cm
\begin{flushleft}
$^1$Department of Physics, Bonner Nuclear Lab, Rice University\\ 6100 Main
Street, Houston, TX 77005, USA\\
$^2$Center for Theoretical Physics, Department of Physics, Texas A\&M
University\\ College Station, TX 77843--4242, USA\\
$^3$Astroparticle Physics Group, Houston Advanced Research Center (HARC)\\
The Mitchell Campus, The Woodlands, TX 77381, USA\\
\end{flushleft}
\end{center}

\vglue 0.5cm
\begin{abstract}
We show that the presence of an anomalous $\rm U_A(1)$ factor in the gauge
group of string-derived models may have the new and important phenomenological
consequence of allowing the vanishing of ${\rm Str}\,{\cal M}^2$ in the
``shifted" vacuum, that results in the process of cancelling the anomalous
$\rm U_A(1)$. The feasibility of this effect seems to be enhanced by a
vanishing vacuum energy, and by a ``small" value of ${\rm Str}\,{\cal M}^2$
in the original vacuum. In the class of free-fermionic models with vanishing
vacuum energy that we focus on, a necessary condition for this mechanism to be
effective is that ${\rm Str}\,{\cal M}^2>0$ in the original vacuum. A vanishing
${\rm Str}\,{\cal M}^2$ ameliorates the cosmological constant problem and is a
necessary element in the stability of the no-scale mechanism.
\end{abstract}
\vspace{0.1cm}
\begin{flushleft}
\baselineskip=12pt
CTP-TAMU-37/95\\
DOE/ER/40717--16\\
ACT-12/95\\
September 1995
\end{flushleft}

\vfill\eject
\setcounter{page}{1}
\pagestyle{plain}
\baselineskip=14pt

Many realistic string models built to date, especially all those constructed
within the free-fermionic formulation, contain U(1) factors in their gauge
groups with non-vanishing traces. These so-called anomalous U(1)'s may be
understood as the result of truncating the string spectrum to the massless
sector (over which the traces are taken), and do not imply an anomaly in the
full string model. Nonetheless, the low-energy effective theory takes notice
of this effect in the form of a Fayet-Iliopoulos contribution to the D-term
of the anomalous $\rm U_A(1)$. This contribution can be calculated by examining
the conditions under which the anomaly will cancel in the full theory, and is
given by a one-loop string calculation \cite{DSW}
\begin{equation}
{\rm D_A}\to {\rm D_A}+\epsilon\,;\qquad
{\rm D_A}=\sum_i q^i_A|\phi_i|^2\,,\qquad
\epsilon={g^2M^2\over192\pi^2}\,{\rm Tr\,U_A}\ ,
\label{eq:DA}
\end{equation}
where $q^i_A$ is the charge of the $\phi_i$ field under $\rm U_A(1)$, and
$M\approx10^{18}\GeV$ is the relevant mass scale (or reduced Planck mass).
The non-zero shift in $\rm D_A$, if not compensated, has the dire consequence
of breaking supersymmetry at the Planck scale. Interestingly enough, in all
instances where anomalous U(1)'s have been reported in consistent models, it
has been always possible to give vacuum expectation values (vevs) to some
scalar fields charged under $\rm U_A(1)$, such that ${\rm D_A}+\epsilon$
vanishes at the nearby vacuum determined by these vevs. This shifted vacuum
is ``nearby" because typically $\vev{\phi}\sim\sqrt{\epsilon}={\cal
O}({1\over10}M)$.

The vacuum shift needed to cancel the anomalous U(1) is not without
consequence. Indeed, one now has to consider carefully other pieces of the
scalar potential that in the absence of $\rm U_A(1)$ would have vanished
automatically, with all vevs equal to zero. These are the usual F- and
D-flatness conditions
\begin{equation}
\VEV{{\partial W\over\partial\phi_i}}=0\,,\qquad
\vev{D_a}=\sum_i q_a^i\,|\vev{\phi_i}|^2=0\ ,
\label{eq:DF}
\end{equation}
where the $a$ label runs over all non-anomalous U(1) factors in the gauge
group. In practice, several of the U(1) factors may have non-zero traces, and
one needs to find the linear combination which is truly anomalous, leaving
the remaining orthogonal linear combinations traceless. This anomalous
combination is given by \cite{HFM}
\begin{equation}
{\rm U_A}={1\over{\rm Tr\,U_A}}\sum_i[{\rm Tr\,U}_i]\,{\rm U}_i\ ,
\label{eq:UA}
\end{equation}
with ${\rm Tr\,U_A}=\sum_i\, [{\rm Tr\,U}_i]^2$. The result of this exercise is
the restoration of supersymmetry in the shifted vacuum, which is not fully
specified, as typically there are many more scalar fields capable of obtaining
vevs than constraint equations restricting the values of these vevs.  Moreover,
this mechanism has an unsuspected phenomenological benefit, as the ratio
\cite{KLN}
\begin{equation}
{\vev{\phi}\over M}\sim{\sqrt{\epsilon}\over M}\sim{1\over10}\ ,
\label{eq:ratio}
\end{equation}
can be used to generate hierarchies in the fermion mass matrices, when the
corresponding Yukawa couplings appear at the non-renormalizable level (see \eg,
Ref.~\cite{decisive}).
\newpage

In this note we would like to point out a new and important consequence of the
vacuum shifting required to cancel the anomalous $\rm U_A(1)$. This pertains
to the calculation of the quantity ${\rm Str}\,{\cal M}^2$ in spontaneously
broken supergravity models, as those obtained in the string models mentioned
above. This quantity is particularly important because it parametrizes a
one-loop quadratic divergence in the scalar potential \cite{FKZ}
\begin{equation}
{1\over32\pi^2}\,{\rm Str}\,{\cal M}^2\, M^2_{\rm Pl}\,,\qquad {\rm Str}\,{\cal
M}^2=2Qm^2_{3/2}\ ,
\label{eq:quadratic}
\end{equation}
where the second expression defines the quantity $Q$ and makes explicit the
dependence on the supersymmetry-breaking order-parameter $m_{3/2}$. After
supersymmetry breaking, if $Q\not=0$ one would generate a cosmological constant
at high scales, which is unlikely to be cancelled by lower energy effects.
This should be motivation enough to seek models with vanishing values of $Q$.
Yet, there are further reasons to desire a suppressed value of $Q$. The
gravitino mass may remain as an undetermined parameter down to low energies, as
advocated in the context of no-scale supergravity
\cite{Cremmer,EKNI+II,LNreview}, where the tree-level cosmological constant is
naturally zero ($V_0=0$). If the condition ${\rm Str}\,{\cal M}^2=0$ is
satisfied \cite{EKNI+II}, then the scalar potential does not depend on large
mass scales, and lower energy dynamical effects may lead to the determination
of the gravitino mass via the no-scale mechanism \cite{Lahanas}. The no-scale
supergravity program has been recently extended to string models \cite{FKZ,LN},
where the most pressing question has become the search for models or mechanisms
by which $Q$ may be sufficiently suppressed.

Ignoring the effect of the anomalous $\rm U_A(1)$, calculations of $Q$ in a few
string models exist. Except for the ``toy" models considered in
Ref.~\cite{FKZ}, no realistic model has yet been found where $Q$ vanishes.
Whatever effects the cancellation of $\rm U_A(1)$ may have, these will probably
be ``small", simply because the vacuum is shifted to a nearby one. In fact,
the non-trivial dimensionless numbers one can form with the vevs are
proportional to the small ratio in Eq.~(\ref{eq:ratio}). As we will see, this
expectation is borne out in specific model-dependent calculations. Therefore,
if $Q$ itself may be shifted towards zero as a consequence of the vacuum shift,
$Q_0$ (the initial value of $Q$) better be ``close" to zero to begin with. The
known model with $Q_0$ closest to zero is that derived in Ref.~\cite{search},
which gives $Q_0=4$ (and $V_0=0$) \cite{LN}. Other known calculations of
$|Q_0|$ yield much larger values, perhaps not unrelated to the fact that
$V_0\not=0$ in those models.

The calculation of $Q$ can be performed in two complementary ways: (i) one can
employ an explicit formula that depends on the K\"ahler function
($G=K+\ln|W|^2$) and the gauge kinetic function, or (ii) one can compute
the supersymmetry-breaking spectrum explicitly and then calculate ${\rm
Str}\,{\cal M}^2$ directly. Here we follow the second approach, as we find it
physically more intuitive. For concreteness, we will restrict our explicit
calculations to free-fermionic string models, although we expect that similar
effects should occur in other string constructions. Furthermore, we will
consider the class of (level-one) free-fermionic models with vanishing vacuum
energy. In such class of models the K\"ahler potential is generically given by
\cite{LN}
\begin{eqnarray}
K&=&-\ln(S+\bar S)
-\ln\left[(\tau+\bar
\tau)^2-\sum_i^{n_{U_1}}[\alpha^{(1)}_i+\bar\alpha^{(1)}_i]^2\right]
+\sum_i^{n_{U_2}}\alpha^{(2)}_i\bar\alpha^{(2)}_i
+\sum_i^{n_{U_3}}\alpha^{(3)}_i\bar\alpha^{(3)}_i\nonumber\\
+&&\!\!\!\!\!\!\sum_i^{n_{T_1}}\beta^{(1)}_i\bar\beta^{(1)}_i
+{1\over\left[(\tau+\bar
\tau)^2-\sum_i^{n_{U_1}}[\alpha^{(1)}_i+\bar\alpha^{(1)}_i]^2\right]^{1/2}}
\left(\sum_i^{n_{T_2}}\beta^{(2)}_i\bar\beta^{(2)}_i
+\sum_i^{n_{T_3}}\beta^{(3)}_i\bar\beta^{(3)}_i\right)
\label{eq:completeK}
\end{eqnarray}
where $S$ and $\tau$ represent the dilaton and modulus fields,
$\alpha^{(1,2,3)}_i,\beta^{(1,2,3)}_i$ represent untwisted and twisted matter
fields in each of the three ``sets" into which the spectrum divides itself
in this class of models. The number of untwisted and twisted fields in each of
these sets is represented by $n_{U_{1,2,3}}$, $n_{T_{1,2,3}}$, and vary from
model to model. Equation~(\ref{eq:completeK}) indicates that the modulus field
$\tau$ corresponds to a field in the first untwisted set. If we ignore the
possible presence of the anomalous $\rm U_A(1)$, the (tree-level) vacuum energy
can be readily shown to vanish ($V_0=0$) for the K\"ahler potential in
Eq.~(\ref{eq:completeK}) \cite{LN}. As indicated above, we calculate
${\rm Str}\,{\cal M}^2=\sum_j(-1)^{2j}(2j+1){\cal M}^2_j$ directly. With the
explicit form of $K$ above, we can calculate each of the mass matrices
separately \cite{LN}. The masses of the complex scalars ($j=0$) are given by
the following multiples of $m_{3/2}$
\begin{equation}
\begin{tabular}{ccrcccr}
$U^{(1)}\,$:&$\alpha^{(1)}_i$& $0$ &\qquad\qquad &$T^{(1)}\,$:&$\beta^{(1)}_i$&
$1$\\
$U^{(2)}\,$:&$\alpha^{(2)}_i$& $1$ & &$T^{(2)}\,$:&$\beta^{(2)}_i$& $0$\\
$U^{(3)}\,$:&$\alpha^{(3)}_i$& $1$ & &$T^{(3)}\,$:&$\beta^{(3)}_i$& $0$\\
\end{tabular}
\label{eq:ScalarMasses}
\end{equation}
and contribute to the supertrace (in units of $m^2_{3/2}$) in the amount of
$2(n_{U_2}+n_{U_3}+n_{T_1})$. The masses of the Majorana fermions ($j=1/2$) are
given by
\begin{equation}
m_{\alpha^{(2)}_i}=m_{\alpha^{(3)}_i}=0\ ,\qquad
m_{\beta^{(1)}_i}=m_{\beta^{(2)}_i}=
m_{\beta^{(3)}_i}=0\ ,
\label{eq:mfermions0}
\end{equation}
and by the eigenvalues of the following mass matrix
\begin{equation}
\left(M_f\right)_{IJ}=m_{3/2}\ \bordermatrix{
& S&\tau&\alpha^{(1)}_j\cr
 S&2/3&-\sqrt{2}/3&0\cr
\tau&-\sqrt{2}/3&1/3&0\cr
\alpha^{(1)}_i&0&0&\delta_{ij}\cr}\ .
\label{eq:FermionMassMatrix}
\end{equation}
That is
\begin{equation}
m_{\eta_\perp}=m_{3/2}\ ,\qquad m_{\alpha^{(1)}_i}=m_{3/2}\ ,
\label{eq:mfermions1}
\end{equation}
where $\eta_\perp$ is the state orthogonal to the massless goldstino (given
by $\eta\propto S+\sqrt{2}\,\tau$). Thus, the Majorana fermions contribute
$-2(1+n_{U_1})$ to ${\rm Str}\,{\cal M}^2$. The Majorana gaugino masses are all
equal to $m_{3/2}$ and contribute $-2d_f$, where $d_f$ is the
dimension of the gauge group.  Finally the gravitino contributes $-4$. Putting
it all together gives
\begin{equation}
Q_0=n_{U_2}+n_{U_3}+n_{T_1}-n_{U_1}-d_f-3\ .
\label{eq:Q0}
\end{equation}
This expression for $Q_0$ may be positive or negative, and will be ``small"
if there is a definite correlation among the numbers of fields in the different
sets and the dimension of the gauge group. This form of the expression (\ie,
the relative signs of the various terms) depends on the number of moduli, which
in turn determine the vacuum energy ($V_0$). In specific examples (not
considered here) one can see that if $V_0\not=0$, then the relative signs in
$Q_0$ do not favor a large cancellation among the various contributions,
resulting in large values of $|Q_0|$ \cite{LN}. If $V_0=0$, the signs are well
balanced and a small value of $Q_0$ is possible. In the only known model of
this class where $V_0=0$ \cite{search},\footnote{It is amusing to note that the
model of Ref.~\cite{search} contains extra matter representations
(\r{10},\rb{10}) sufficient to postpone the gauge coupling unification scale
naturally up to the string scale $M$.} one has
$n_{U_1}=13,n_{U_2}=14,n_{U_3}=16$; $n_{T_1}=80,n_{T_2}=80,n_{T_3}=68$;
and $d_f=90$, and therefore
\begin{equation}
Q_0=14+16+80-13-90-3=110-106=4\ .
\label{eq:Q0ex}
\end{equation}
We say $Q_0=4$ is ``small" in the sense that it is only 2\% of the total
obtained if the terms had been added in absolute value.

When one takes into account the presence of the anomalous $\rm U_A$,
Eqs.~(\ref{eq:DA},\ref{eq:DF}) need to be satisfied in non-trivial ways.
The possible vacuum expectation values of the scalar fields are further
constrained by the desire to maintain $V_0=0$. This can be assured by
allowing non-zero vevs only for the scalar fields that do not acquire
supersymmetry-breaking masses, \ie, those in the $U^{(1)}$, $T^{(2)}$, and
$T^{(3)}$ sets (see Eq.~(\ref{eq:ScalarMasses})):
\begin{equation}
\VEV{\alpha^{(2)}_i}=\VEV{\alpha^{(3)}_i}=\VEV{\beta^{(1)}_i}=0\ .
\label{eq:zero-vevs}
\end{equation}
To determine $Q$ in the presence of non-zero vevs, we need to revisit the
contributions to ${\rm Str}\,{\cal M}^2$. The scalar masses are not affected
by shifts in vevs.\footnote{This is true at tree-level. The one-loop string
effect that gives the Fayet-Iliopoulos contribution to $\rm D_A$, induces mass
shifts $\sim\sqrt{\epsilon}$ for all scalars charged under $\rm U_A(1)$. These
mass shifts do not contribute to ${\rm Str}\,{\cal M}^2$ since, when $\rm D_A$
is cancelled, compensating one-loop fermionic mass shifts are generated, such
that supersymmetry is restored in the shifted vacuum \cite{DL}.} The gaugino
masses are also unaffected, as (at tree-level) they only depend on the dilaton
contribution to the gauge kinetic function. The gravitino contribution is also
unaffected since $m_{3/2}$ is an overall factor. All we need to consider are
the Majorana fermion masses, which are given by \cite{FKZ}
\begin{equation}
\left(M_f\right)_{IJ}=m_{3/2}\left(G_{IJ}-G_{IJ\bar K}G^{\bar
K}+\coeff{1}{3}G_IG_J\right)\ .
\label{eq:FermionMasses}
\end{equation}
Examining the K\"ahler potential in Eq.~(\ref{eq:completeK}), we see that we
still have
\begin{equation}
m_{\alpha^{(2)}_i}=m_{\alpha^{(3)}_i}=m_{\beta^{(1)}_i}=0\ ,
\label{eq:mfermions0A}
\end{equation}
whereas the matrix in Eq.~(\ref{eq:FermionMassMatrix}) is extended to also
include the $\beta^{(2)}_i,\beta^{(3)}_i$ fields. In what follows we will make
the simplifying assumption that the vev shifts are small compared to the vev of
the modulus field: $\vev{\alpha,\beta}/\vev{\tau}\ll1$. This assumption is
motivated by the fact that one would expect $\vev{\tau}\sim M$, whereas the
anomalous $\rm U_A(1)$ cancellation generically implies $\vev{\alpha,\beta}\sim
{1\over10}M$. (Note that the modulus field is a gauge singlet and does not
participate in the $\rm U_A(1)$ cancellation mechanism.) Should this
expectation not be realized, the following calculations would need to be
performed numerically. The resulting symmetric matrix of properly normalized
fields is given by
\begin{equation}
\left(M_f\right)_{IJ}=m_{3/2}\ \bordermatrix{
& S&\tau&\alpha_1^{(1)}&\alpha_2^{(1)}&\beta_1^{(2,3)}&\beta_2^{(2,3)}\cr
 S&{2\over3}&-{\sqrt{2}\over3}(1+X+{1\over2}Y)&{\sqrt{2}\over3}\sqrt{X_1}
&{\sqrt{2}\over3}\sqrt{X_2}&{1\over3}\sqrt{Y_1}&{1\over3}\sqrt{Y_2}\cr
\tau&&{1\over3}(1+5X+4Y)&-{4\over3}\sqrt{X_1}&-{4\over3}\sqrt{X_2}
&{-1\over3\sqrt{2}}\sqrt{Y_1}&{-1\over3\sqrt{2}}\sqrt{Y_2}\cr
\alpha_1^{(1)}&&&1-{2\over3}X_1&\sqrt{X_1X_2}
&\sqrt{X_1Y_1}&\sqrt{X_1Y_2}\cr
\alpha_2^{(1)}&&&&1-{2\over3}X_2&\sqrt{X_2Y_1}&\sqrt{X_2Y_2}\cr
\beta_1^{(2,3)}&&&&&-{1\over3}Y_1&Y_1Y_2\cr
\beta_2^{(2,3)}&&&&&&-{1\over3}Y_2\cr}
\label{eq:FermionMassMatrixA}
\end{equation}
Here we have restricted the number of generic fields to the minimal that show
the emergent pattern. In this matrix we have defined the ratios
\begin{equation}
X_i={\left(\alpha_i^{(1)}+\bar\alpha_i^{(1)}\right)^2\over(\tau+\bar\tau)^2}\,,
\quad Y_i={\beta_i^{(2,3)}\bar\beta_i^{(2,3)}\over\tau+\bar\tau}\ ,
\label{XiYi}
\end{equation}
and the sums of ratios
\begin{equation}
X=\sum_i X_i\ ,\qquad Y=\sum_i Y_i\ ,
\label{XY}
\end{equation}
which run over all the fields in the first untwisted set ($\alpha^{(1)}_i$) and
the second and third twisted sets ($\beta^{(2)}_i,\beta^{(3)}_i$). Also, the
coefficients of the higher-order off-diagonal terms ($\sqrt{X_iY_j}$)
in Eq.~(\ref{eq:FermionMassMatrixA}) have been omitted.
Note that the $X_i$ and $Y_i$ corresponding to fields charged under the
Standard Model quantum numbers vanish on phenomenological grounds. For our
present purposes, it is enough to compute the trace of the square of the
Majorana mass matrix. This amounts to summing over the squares of all of the
elements of the $M_{IJ}$ mass matrix. We obtain
\begin{equation}
{\rm Tr}M^2_{IJ}=m^2_{3/2}
\left[1+n_{U_1}+\coeff{14}{3}X+\coeff{5}{3}Y\right]\ ,
\label{eq:Str}
\end{equation}
keeping terms to only first order in $X$ and $Y$, since the twisted sector
K\"ahler potential is only known to first order. The expression for $Q$ then
becomes
\begin{equation}
Q=Q_0-\Delta Q=Q_0-\coeff{1}{3}(14X+5Y)\ ,
\label{eq:Qnew}
\end{equation}
with $Q_0$ as given in Eq.~(\ref{eq:Q0}). Since $X$ and $Y$ are both positive,
we conclude that a {\em necessary} condition for a possibly vanishing $Q$ in
the shifted vacuum, is that $Q_0>0$ in the original vacuum. If $Q_0$ is
positive and ``small", then $Q$ may vanish in a suitably chosen shifted
vacuum.

The choice of shifted vacuum is a rather model-dependent exercise, as the
possible choices of vacuum expectation values are restricted by the D- and
F-flatness constraints in Eqs.~(\ref{eq:DA},\ref{eq:DF}), plus the $V_0=0$
constraint in Eq.~(\ref{eq:zero-vevs}). The idea is to scan the parameter
space of simultaneous solutions to these set of equations, looking for
values of the vevs which bring $Q$ closest to zero. Is this a feasible
possibility? At this point we can say that we have carried out this exercise
in the model of Ref.~\cite{search}, which has $Q_0=4$, and have found sets
of values for the vevs that indeed shift $Q_0$ down to zero. Details of this
particular application are given elsewhere \cite{constraints}. We can however
show that this specific case is rather typical. Let us assume that
all vevs have the same magnitude: $\vev{\phi}\sim\sqrt{\epsilon}$, and consider
the expressions for $X$ and $Y$
\begin{equation}
X\sim (n_{U_1})\gamma\ ,
\qquad Y\sim (n_{T_2}+n_{T_3})\sqrt{\gamma}\,{\sqrt{\epsilon}\over M}\ ,
\label{eq:Xapp}
\end{equation}
where we have defined
\begin{equation}
\gamma=\left({\sqrt{\epsilon}\over\tau+\bar\tau}\right)^2\ .
\label{eq:gamma}
\end{equation}
Taking typical values of the number of untwisted and twisted fields in a set
($n_{U_1}\sim10$, $n_{T}\sim80$) we see that
\begin{equation}
\coeff{1}{3}(14X+5Y)\sim \coeff{140}{3}\,\gamma+\coeff{80}{3}\,\sqrt{\gamma}\ ,
\label{eq:app}
\end{equation}
where we have set $\sqrt{\epsilon}/M\sim{1\over10}$. Plugging in numbers one
finds that ${1\over3}(14X+5Y)\sim{\rm few}$ for $\gamma\sim0.01$. Since we
expect $\vev{\tau}\sim M$, and thus $\gamma\sim(\sqrt{\epsilon}/M)^2\sim0.01$,
we see that this mechanism is quite feasible, if $Q_0\sim{\rm few}$. The
explicit model-dependent calculations in Ref.~\cite{constraints} yield similar
results.

Let us point out that restricting the shifted vacua to those where $Q$ nearly
vanishes is a phenomenological procedure. In fact, adding to the scalar
potential  the one-loop contribution in Eq.~(\ref{eq:quadratic}) does not shift
the vevs in any noticeable way, since this contribution ($\sim m^2_{3/2}M^2$)
is so much smaller than that coming from $D_A^2$ ($\sim \epsilon^2\phi^2\sim
M^4$). One may speculate that in the full string model new (loop) contributions
to the scalar potential arise that make the chosen vacuum (where $Q$ vanishes)
energetically preferred. Our phenomenological procedure would then be the
``effective" way of realizing such scenario. These corrections will further
shift the value of $Q$, as the K\"ahler potential and the gauge kinetic
function receive string loop corrections. Thus, it is not clear that at this
level of approximation one should be fixated on an exactly vanishing value of
$Q$. It has been suggested \cite{FKZ}, that in analogy with the string loop
corrections to the gauge kinetic function, which vanish at two and higher
loops, perhaps $Q$ would behave similarly. In this case, the higher-loop
quadratic divergencies pointed out recently in Ref.~\cite{Bagger}, would
effectively vanish in the full string theory.

In sum, we have shown that the presence of an anomalous $\rm U_A(1)$ factor
in the gauge group of string models may have yet another welcomed
phenomenological consequence, since its cancellation may allow the vanishing of
${\rm Str}\,{\cal M}^2$ in the shifted vacuum. This result seems to be the
more likely if the vacuum energy vanishes, and if ${\rm Str}\,{\cal M}^2$
is ``small" in the original vacuum. In the class of free-fermionic
models that we have focused on, ${\rm Str}\,{\cal M}^2$ must also be positive
in the original vacuum. Evidently, having ${\rm Str}\,{\cal M}^2=0$ has
important consequences for the cosmological constant problem, and for the
stability of the no-scale mechanism.

\newpage
\section*{Acknowledgments}
The work of J.~L. has been supported in part by DOE grant DE-FG05-93-ER-40717.
The work of D.V.N. has been supported in part by DOE grant DE-FG05-91-ER-40633.

\end{document}